\documentclass[10pt]{article}

\usepackage{graphicx}
\usepackage{authblk}
\usepackage{multirow}
\usepackage[fleqn]{amsmath}
\usepackage[margin=3.5cm]{geometry}
\usepackage{natbib}
\usepackage{lineno}
\usepackage{float}

\def \hat {\widehat}
\def \bbeta{\mbox{\boldmath$\beta$}}

\def \btheta{\mbox{\boldmath$\theta$}}

\def \bmu{\mbox{\boldmath$\mu$}}
\def \bsigma{\mbox{\boldmath$\sigma$}}

\def \brho{\mbox{\boldmath$\rho$}}

\def \bY{\mathbf{Y}}
\def \bX{\mathbf{X}}

\def \b1{\mathbf{1}}
\def \bY{\mathbf{Y}}

\title{\large{\textbf{Modeling data with zero inflation and overdispersion using GAMLSSs}}  }

\author[1]{{\normalsize{Gustavo Thomas}}\thanks{Corresponding author. E-mail: gustavothomas17@usp.br}}
\author[2]{{\normalsize{Luiz R. Nakamura}}}
\author[1]{{\normalsize{Rafael A. Moral}}}
\author[1]{{\normalsize{Clarice G.B. Dem\'{e}trio}}}
\affil[1]{{\small Departamento de Ci\^{e}ncias Exatas, Escola Superior de Agricultura Luiz de Queiroz, Universidade de S\~{a}o Paulo, Av. P\'{a}dua Dias, 11, Piracicaba, SP, Brazil}}
\affil[2]{{\small Departamento de Inform\'{a}tica e Estat\'{i}stica, Universidade Federal de Santa Catarina, Campus Universit\'{a}rio Trindade, Florian\'{o}polis, SC, Brazil}}

\date{}

\begin{document}

\maketitle

\section*{Abstract}
Count data with high frequencies of zeros are found in many areas, specially in biology. Statistical models to analyze such data started being developed in the 80s and are still a topic of active research. Such models usually assume a response distribution that belongs to the exponential family of distributions and the analysis is performed under the generalized linear models' framework. However, the generalized additive models for location, scale and shape (GAMLSSs) represent a more general class of univariate models that can also be used to model zero inflated data. In this paper, the analysis of a data set with excess of zeros and overdispersion already analyzed in the literature is described using GAMLSSs. The specific GAMLSSs' tools used enhanced model comparison and eased the interpretation of results.

\noindent \textbf{Keywords:} Zero inflated models; GAMLSSs; \textit{Trajan} data set

\section{Introduction}

When analysing count data, it is common to encounter a higher number of zero responses than predicted by the Poisson distribution. In a biological context, this is often a result of individuals that respond in two distinct ways to the effect of interest. While one group of individuals is not able to respond and these will always result in zero counts; also known as structural zeros, another might respond or not and occasional zeros are known as sampling zeros. This situation can be described by a two-component mixture model with zero inflation.

The literature of zero-inflated models dates back to the 80s with \citet{Mullahy86}, who was the first to define a two-part model. Six years later, \citet{Lambert92} introduced the zero-inflated Poisson (ZIP) model with an application to manufacturing defects. These ideas originated a whole class of new models such as the zero-inflated binomial (ZIB) model, the zero-inflated negative binomial (ZINB) model \citep{Greene94}, zero-altered models  \citep{Heilbron94}, mixed versions of the ZIP and ZIB \citep{Hall00}, among others.

\citet{Ridout98} make a review of the literature on ZIP and other closely related zero-inflated models for unbounded counts and present a comparison of these models for an horticulture experiment (\textit{Trajan} data) using a generalized linear model (GLM) \citep{Nelder72} approach. On that occasion, they found that the ZINB model provided the best fit over the traditional Poisson, negative binomial and ZIP models, comparing different predictors for each distribution parameter.

This work describes the modeling of the \textit{Trajan} data using ZIP, ZINB, ZIPIG (zero-inflated Poisson inverse Gaussian) and ZIBNB (zero-inflated beta negative binomial) distributions using the generalized additive models for location, scale and shape (GAMLSSs) \citep{Rigby05} framework instead of the GLM approach. Specific GAMLSSs tools are used for the comparison of models and adequacy checking.

\section{Trajan data}

The data recorded in the horticulture experiment were the number of roots produced by 270 micropropagated shoots of the columnar apple cultivar \textit{Trajan} \citep{Ridout98}. All shoots were kept under identical conditions apart from concentrations of a hormone and photoperiod durations inside growth cabinets. Two growth cabinets were used, one that received 8 and the other 16 hours of photoperiod per day. Jars with one shoot each were placed at random in each of the two cabinets. In addition, four concentrations of the cytokinin BAP hormone (2.2, 4.4, 8.8 and 17.6$\mu$M) were used, so that the experiment had a 4$\times$2 factorial design of treatments. The objective of the study was to evaluate the effects of the two factors (concentration of cytokinin and photoperiod duration) on the number of roots produced. The full description of the experiment can be found in \citet{Marin93}. 

\begin{figure}[!htb]
	\centering
	\includegraphics[width=10cm]{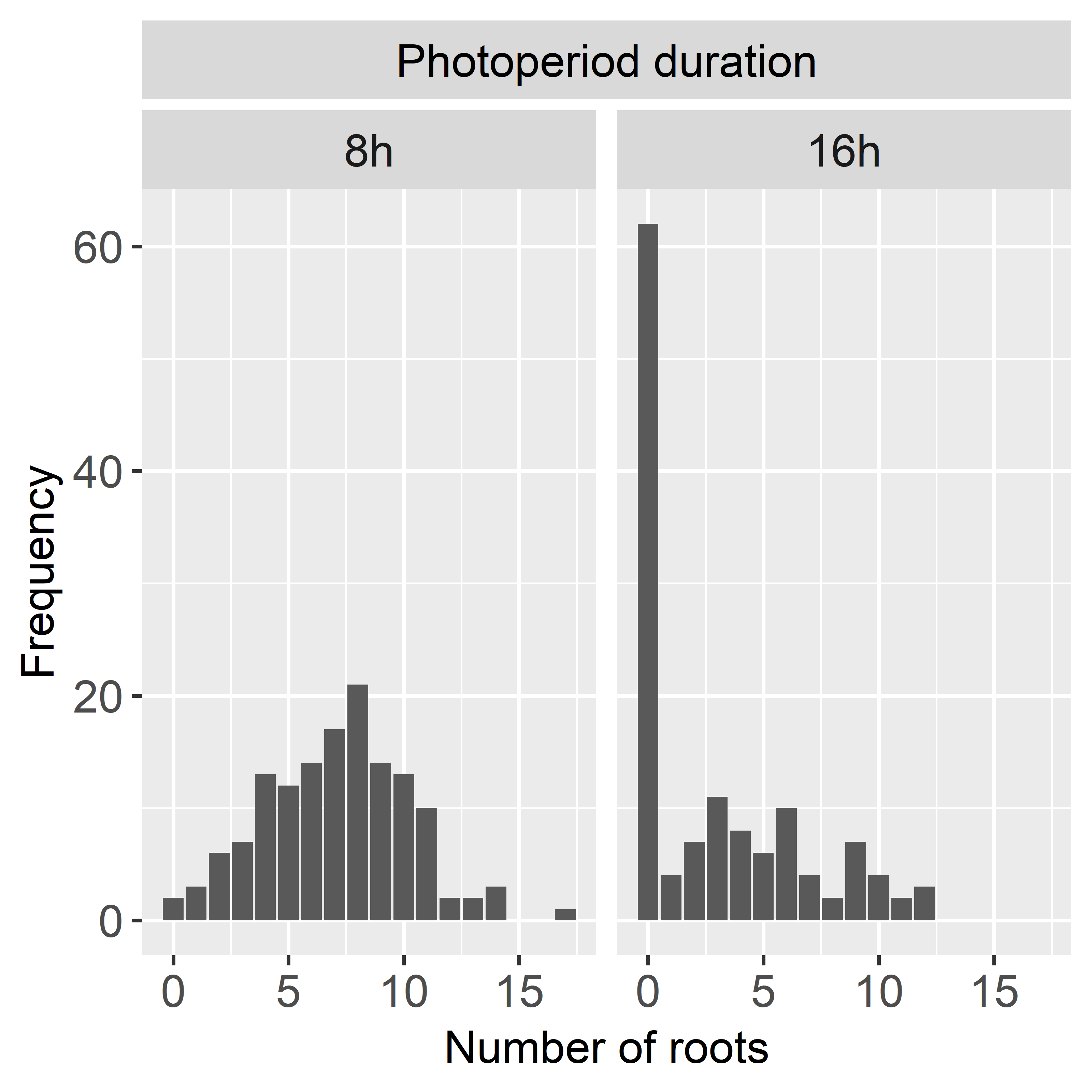}
	\caption{Barplots of the number of roots per each photoperiod duration.}
	\label{bps}
\end{figure}

The key feature of this data set is that almost half of the shoots under the 16-hour photoperiod did not root, as seen in Table \ref{data} and Figure \ref{bps}. This means that a zero-inflated model that accounts for this excess of zeros is needed. \citet{Ridout98} pointed out that the data are also overdispersed, which can also be seen from the two bottom lines of mean and variance in Table \ref{data}. The variances of the number of roots generated in the cabinet with 16 hours of photoperiod were more than three times higher than the corresponding means for all benzo(a)pyrene (BAP) concentrations, while for the other treatment combinations this ratio is much smaller, see Table \ref{data}.	

\begin{table}[H]
	\centering
	\caption{\textit{Trajan} data set: number of shoots that rooted by number of roots and treatments.}
	\label{data}
	\begin{tabular}{cllllllll}
		\cline{1-9}
		& \multicolumn{8}{c}{Photoperiod}                                          \\ 
		& \multicolumn{4}{c}{8 hours} & \multicolumn{4}{c}{16 hours}               \\ \cline{2-9} 
		\multicolumn{1}{l}{BAP ($\mu$M)}     & 2.2 & 4.4 & 8.8 & 17.6 & 2.2 & 4.4 & 8.8 & \multicolumn{1}{c}{17.6} \\ \cline{2-9} 
		\multicolumn{1}{l}{Number of roots}  &      &      &      &      &     &     &     &                          \\
		0                                 & 0     & 0    & 0    & 2     & 15  & 16  & 12  & 19                       \\
		1                                 & 3     & 0    & 0    & 0     & 0   & 2   & 3   & 2                        \\
		2                                 & 2     & 3    & 1    & 0     & 2   & 1   & 2   & 2                        \\
		3                                 & 3     & 0    & 2    & 2     & 2   & 1   & 1   & 4                        \\
		4                                 & 6     & 1    & 4    & 2     & 1   & 2   & 2   & 3                        \\
		5                                 & 3     & 0    & 4    & 5     & 2   & 1   & 2   & 1                        \\
		6                                 & 2     & 3    & 4    & 5     & 1   & 2   & 3   & 4                        \\
		7                                 & 2     & 7    & 4    & 4     & 0   & 0   & 1   & 3                        \\
		8                                 & 3     & 3    & 7    & 8     & 1   & 1   & 0   & 0                        \\
		9                                 & 1     & 5    & 5    & 3     & 3   & 0   & 2   & 2                        \\
		10                        & 2     & 3    & 4    & 4     & 1   & 3   & 0   & 0                        \\ 
		11                        & 1     & 4    & 1    & 4     & 1   & 0   & 1   & 0                        \\ 
		12                        & 0     & 0    & 2    & 0     & 1   & 1   & 1   & 0                        \\ 
		13                        & 1     & 1    & 0    & 0     & 0   & 0   & 0   & 0                        \\ 
		14                        & 0     & 0    & 2    & 1     & 0   & 0   & 0   & 0                        \\ 
		17                        & 1     & 0    & 0    & 0     & 0   & 0   & 0   & 0                        \\ \hline 
		\multicolumn{1}{l}{Number of shoots} & 30    & 30   & 40   & 40    & 30  & 30  & 30  & 40                       \\ 
		\multicolumn{1}{l}{Mean} & 5.8 & 7.8& 7.5& 7.2& 3.3& 2.7& 3.1& 2.5    \\ 
		\multicolumn{1}{l}{Variance}&14.1&7.6&8.5&8.8&16.6&14.8&13.5&8.5 \\ \hline
		{Source: \citet{Marin93}.}
	\end{tabular}
\end{table}

\section{Modeling}

\subsection{Generalized additive models for location, scale and shape (GAMLSSs)}

In this paper, the \textit{Trajan} data set is analyzed using GAMLSSs. These type of models allow any distribution to be assumed for the response, not restricting to the ones belonging to the exponential family of distributions. Furthermore, under this framework all parameters of the distribution can be modeled explicitly by predictors. The GAMLSSs generalize not only the GLMs but also the generalized additive models (GAMs) of \citet{Hastie90} in the sense that parameters are linked to predictors through known monotonic link functions that can include a wide range of linear/non-linear terms, as well as random effects and nonparametric terms. In this application, though, only linear functions of the predictors were used, so that the underlying model is a parametric GAMLSS.

A parametric GAMLSS assumes that an independent random variable $Y_i,i=1,...,n$ follows a probability distribution $\mathbf{D}_{\btheta}$. For simplicity, suppose that $\mathbf{D}_{\btheta}$ has four parameters so that $\btheta$=($\theta_{1},\theta_{2},\theta_{3},\theta_{4}$)'=($\mu,\sigma,\nu, \tau$)'. Parameters of the distribution are linked to linear predictors by link functions the same way as in generalized models, but in the GAMLSS's framework this is made for every parameter of the distribution, not restricted to the mean of the distribution (commonly its first parameter). Distribution parameters are conventionally denoted by $\mu$, $\sigma$, $\nu$ and $\tau$ (in this order) in the context of GAMLSSs. For most of the distributions implemented for GAMLSSs \citep{Stasinopoulos07}, the first parameter $\mu$ is the location parameter associated with the mean of the distribution and $\sigma$ is a scale parameter associated with the variance. The remaining parameters are usually related to the shape of the distribution. 

For example, suppose that the number of roots in the experiment described follows a (fictional) distribution $\mathbf{D}$ with two parameters: $\mu$ and $\sigma$. Assume that the mean number of roots generated (represented by the $\mu$ parameter) is modeled by the main effects of photoperiod and BAP concentration factors and the zero inflation parameter (represented by the second parameter $\sigma$) is allowed to depend on the photoperiod duration only. The corresponding model would be represented as 

\begin{equation}
\begin{array}{c}
\bY \sim \mathbf{D}(\bmu,\bsigma) \\

g_1(\bmu) =  \bX_1 \brho + \bX_2 \bbeta \\

g_2(\bsigma) =  \bX_1 \brho\mbox{'}, 

\end{array}
\end{equation}
where $\bX_1$ and $\bX_2$ denote the design matrices of the photoperiod durations and BAP concentrarions respectively. The vectors $\brho$ and $\brho$' contain the parameters for the photoperiod concentrations when modeling $\mu$ and $\sigma$, respectively, and $\bbeta = (\beta_{2.2},\beta_{4.4},\beta_{8.8},\beta_{17.6})$' represents the BAP concentration parameters. 

The selection procedure of a parametric GAMLSS consists in comparing models with different combinations of distributions for the response variable and predictor terms for each parameter predictor. This means that a suitable zero-inflated distribution will have to be selected to build a GAMLSS for the number of roots in the \textit{Trajan} micropropagation experiment. Zero-adjusted models can also handle higher frequencies of zeros in the response, but since they are two-part models they do not have a good interpretation for this biological process in particular. Such models assume that there is a threshold that increases the number of observed zero counts, which are modeled independently from the process that generates positive counts. Nevertheless, the rooting process of \textit{Trajan} shoots has a higher probability of zeros due to the natural unrooted shoots (sampling zeros) plus another mechanism that would only generate zero responses (structural zeros), which is the idea of a zero-inflated model. \citet{Ridout98} used the Poisson, negative binomial, ZIP and ZINB models in their analysis of the \textit{Trajan} data set. They used the first two of them more as a basis for comparison, since the excess of zeros was clear. The zero-inflated Poisson accounted for that, but did not handle the overdispersion. Hence, the ZINB model provided the best fit to the data in their analysis.

\begin{figure}[!htb]
	\centering
	\includegraphics[width=13cm, height = 12cm]{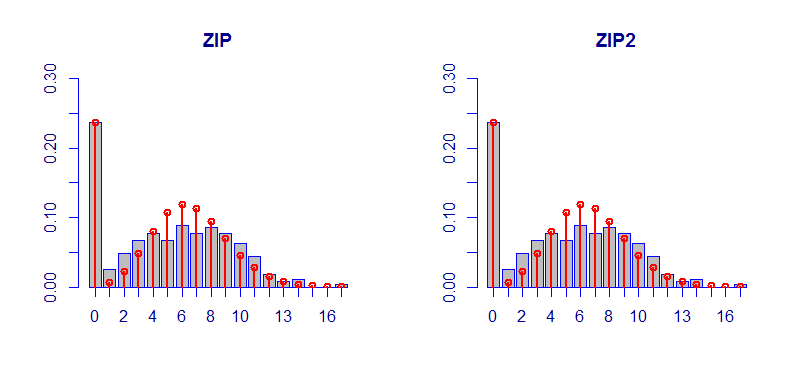}
		\caption{Histograms of the number of roots with fits of selected distributions (PO: Poisson, NB: Negative Binomial, ZIP: Zero Inflated Poisson, ZINB: Zero Inflated Negative Binomial, ZIPIG: Zero Inflated Poisson Inverse Gaussian and ZIBNB: Zero Inflated Beta Negative Binomial).}
	\label{bps2}
\end{figure}

To aid in the procedure of choosing a suitable distribution for a GAMLSS, the \verb|gamlss| package in \verb|R| provides the \verb|histDist()| function. This function produces an histogram of the data set and superimposes a model fit using a response distribution to be tested. It gives an initial idea of the suitability of the distribution to the response variable, since the generated fit is entirely based on the response variable's histogram, not taking into account any covariates. Figure \ref{bps2} shows the plots produced with six distributions using the \verb|histDist()| function for the \textit{Trajan} data set. The distribution fits are represented by the vertical lines with circles at the top. Both the Poisson and negative binomial models (PO and NB) do not account for the zero inflation. Between the ZIP, ZINB, ZIPIG and ZIBNB model fits, however, the only visible difference is that the ZIP overestimates the number of roots a little more than the others, but all four models appear to fit the data reasonably well. Therefore, these four distributions represent good candidates to compose the GAMLSS for the number of roots, and each one of them is better described in the following section.

\subsection{Zero inflated models for GAMLSSs} 

The ZIP distribution used in this analysis is a reparameterized version of the original ZIP model of \citet{Lambert92}, where the first parameter $\mu$ is the mean of the distribution. This is the case for most of the distributions implemented for GAMLSSs and facilitates the interpretation of the model. Similar to the original ZIP, this modified version of the ZIP model also assumes that the response variable has a high frequency of zeros (represented by the zero inflation parameter $\sigma$), and follows a Poisson process for positive counts. Its probability mass function (pmf) is given by

\begin{equation}
P(Y=y|\mu,\sigma) =
\left\{
\begin{array}{ll}
\sigma + (1 - \sigma)e^{-(\frac{\mu}{1-\sigma})}, & \mbox{if } y=0 \\
(1 - \sigma) \frac{\mu^{y}}{y!(1-\sigma)^{y}} e^{-(\frac{\mu}{1-\sigma})}, & \mbox{if } y=1,2,...  \mbox{ ,}\\
\end{array}
\right.
\label{zip}
\end{equation}

and assumes $y=0,1,2, ...$, $\mu \geq 0$, $\sigma \geq 0$ and $0 \leq \nu \leq 1$.
%It can be seen from this equation that the probability of zero counts in this ZIP model is equal to the probability of zeros expected by the Poisson distribution plus the inflation parameter $\sigma$, while the positive counts are modeled by a Poisson process with mean $\mu$. 

The ZINB distribution is similar in form to the ZIP, but it inflates the probability of a zero count by one parameter ($\nu$) and has another ($\sigma$) to adjust the variance independently of the mean. The pmf for the ZINB distribution is equal to
\begin{equation}
P(Y=y|\mu,\sigma,\nu) =
\left\{
\begin{array}{ll}
\nu + (1 - \nu)P(Y=y|\mu,\sigma), & \mbox{if } y=0 \\
(1 - \nu) P(Y=y|\mu,\sigma), & \mbox{if } y=1,2,... ,  \\
\end{array}
\right.
\label{zinb}
\end{equation}
where
\begin{equation}
P(Y=y|\mu,\sigma) = \frac{\Gamma(y+\frac{1}{\sigma})}{\Gamma(\frac{1}{\sigma})
	\Gamma(y+1)}\left(\frac{\sigma \mu}{1+\sigma \mu}\right) ^{y}\left( \frac{1}{1+\sigma \mu}\right) ^{\frac{1}{\sigma}} \\
\label{nbi}
\end{equation}
for $y=0,1,2, ...$, $\mu \geq 0$, $\sigma \geq 0$ and $0 \leq \nu \leq 1$. In fact, equation (\ref{nbi}) is one of the parameterizations of the probability mass function of the negative binomial distribution. 

The Poisson inverse Gaussian is another distribution that has potential for modeling highly dispersed count data due to its flexibility. It is a continously mixed Poisson distribution that allows for even higher skewness (a longer upper tail) than the negative binomial. Its zero inflated version has pmf equal to (\ref{zinb}) with 
\begin{equation}
P(Y=y|\mu,\sigma) = \frac{\mu^{y}e^{1/ \sigma}K_{y-1/2}(\alpha)}{(\alpha \sigma)^{y} y! }\left( \frac{2\alpha}{\pi}\right) ^{\frac{1}{2}}.  \\
\label{pig}
\end{equation}
In (\ref{pig}), $\alpha^2 = \frac{1}{\sigma^2}+\frac{2 \mu}{\sigma}$, $K_{\lambda}(t)=\frac{1}{2} \int_{0}^{\infty}{x^{\lambda-1}\exp[-\frac{1}{2}t(x+x^{-1})]dx}$ (which is known as the modified Bessel function of third kind) and the ZIPIG is only valid for $y=0,1,2, ...,$ $\mu \geq 0$, $\sigma \geq 0$ and $0 \leq \nu \leq 1$. 

Finally, the pmf of the ZIBNB distribution is
\begin{equation}
P(Y=y|\mu,\sigma,\nu) =
\left\{
\begin{array}{ll}
\tau + (1 - \tau)P(Y=y|\mu,\sigma,\nu), & \mbox{if } y=0 \\
(1 - \tau) P(Y=y|\mu,\sigma,\nu), & \mbox{if } y=1,2,... , \\
\end{array}
\right.
\label{zibnb}
\end{equation}
where 
\begin{equation}
P(Y=y|\mu,\sigma,\nu) = \frac{\Gamma(y+\frac{1}{\nu})B(y+\frac{\mu\nu}{\sigma}, \frac{1}{\sigma}+\frac{1}{\nu}+1) }
{\Gamma(y+1)B(\frac{\mu\nu}{\sigma}, \frac{1}{\sigma}+1)}  \\
\label{bnb}
\end{equation}
is a parameterization of the beta negative binomial (BNB) function. The ZIBNB is an overdispersed zero-inflated negative binomial distribution. Its first parameter $\mu$ is the mean count of the supposed independent Bernoulli trials, whose probability of success is assumed to follow a Beta($\sigma$, $\nu$) distribution, whereas $\tau$ is the zero inflation parameter. The next section describes a comparison of GAMLSSs for the \textit{Trajan} data set using these four distributions.

\section{Results and discussion}

An exploratory analysis for the \textit{Trajan} data set is summarized graphically in Figure 3. From the boxplots by treatment shown on the left plot of Figure 3, there is a positive effect of the 8h photoperiod duration on the number of roots and mild effects of the different concentration levels. The interaction plot shown on the right side of Figure 3 gives little evidence of interaction between the factors because the mean lines do not change much in different directions when the BAP concentration increases from 2.2 to 8.8 $\mu$M. In addition, this plot highlights the high frequency of unrooted shoots in the 16h-photoperiod cabinet represented by the big blue circles in the bottom.

\begin{figure}[h]
	\begin{center}
		\begin{minipage}[t]{7cm}
			\includegraphics[width=7cm, height = 7cm]{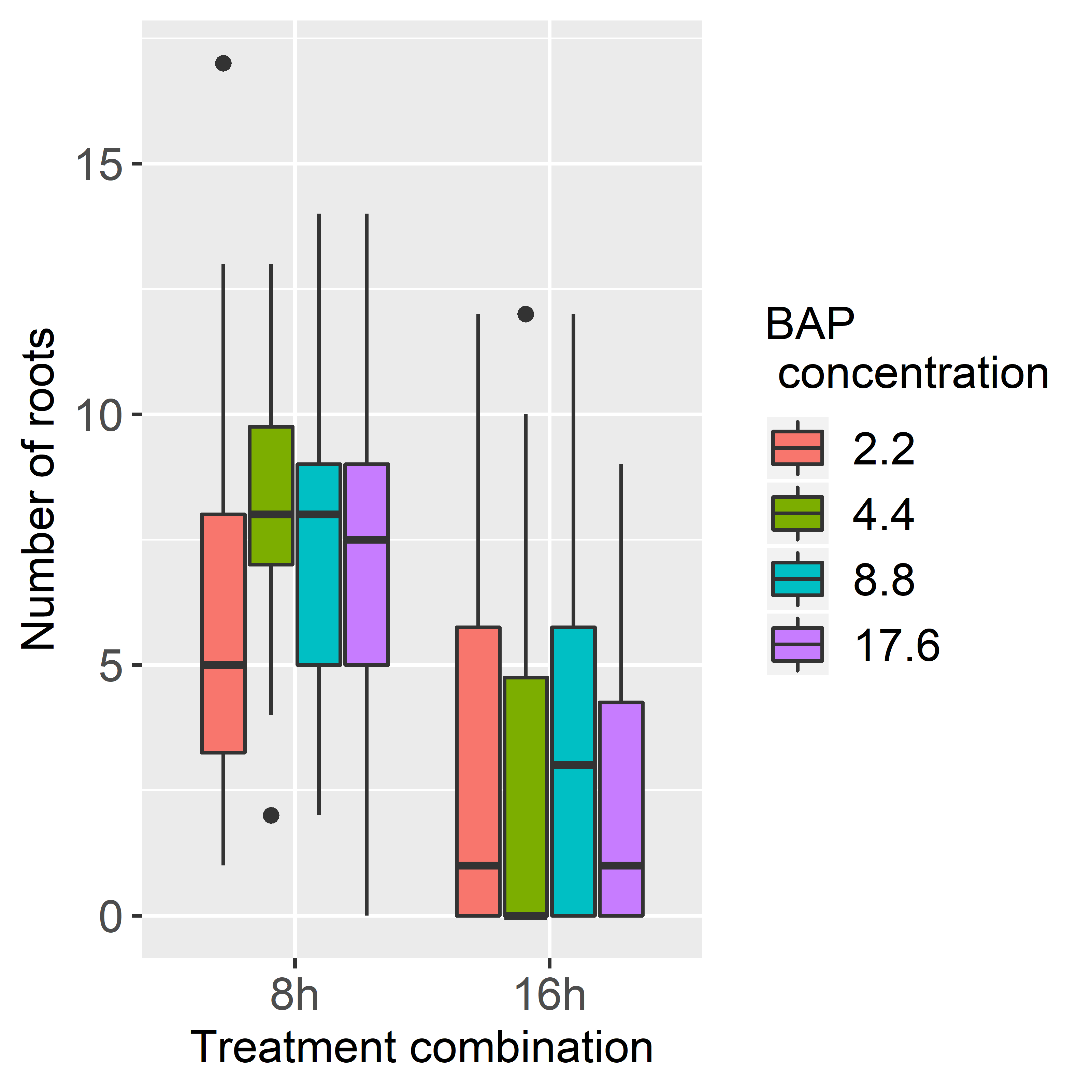}
		\end{minipage}
		\begin{minipage}[t]{7cm}
			\includegraphics[width=7cm, height = 7cm]{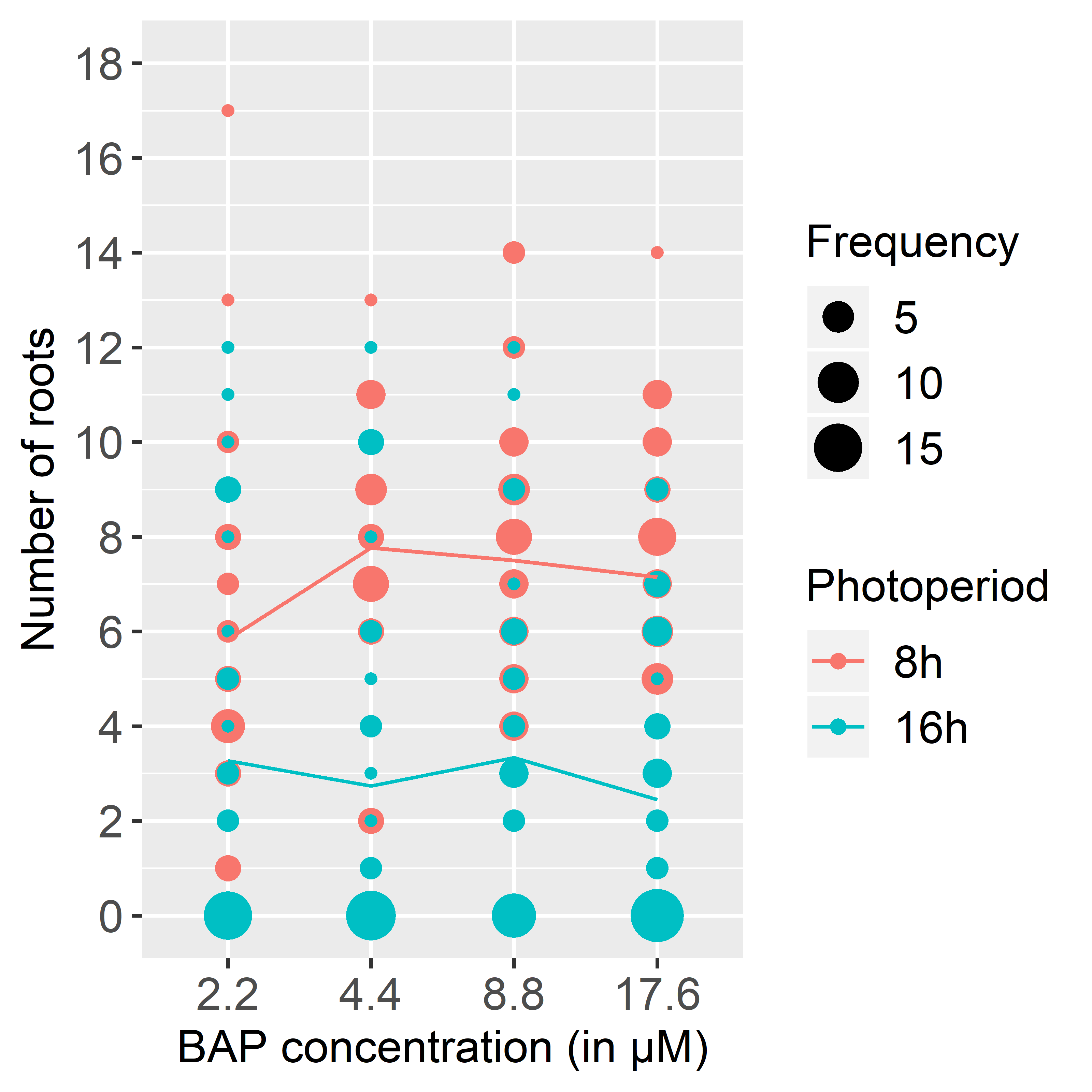}
		\end{minipage}
		\label{box}
	\end{center}
		\caption{Boxplots of each treatment combination for the number of roots in the \textit{Trajan} data set (left). Interaction plot of the factors for the number of roots in the \textit{Trajan} data set. Lines represent the mean number of roots for each photoperiod duration and the frequency of specific number of roots for each treatment combination is given by the size of the circles (right).}
\end{figure}

\subsection{Model selection}

There are several strategies that can be employed to find the best predictor for a parameter of the response variable distribution in a GAMLSS. The usual forward, backward and stepwise procedures are appealing and popularly used. Within the \verb|gamlss| package, the function \verb|stepGAICAll.A()| uses such a procedure with the Akaike information (AIC, \citet{Akaike83}) as the criterion for selection. For a response distribution with set of parameters $(\mu,\sigma,\nu,\tau)^T$, this function's strategy is described in the following sequential steps \citep{Nakamura2017}:
\begin{enumerate}
	\setlength\itemsep{-0.3em}
	\item Use a forward AIC selection procedure to select an appropriate model for $\mu$, with $\sigma$, $\nu$  and $\tau$ fitted as constants;
	
	\item Given the model for $\mu$ obtained in 1, and for $\nu$ and $\tau$ fitted as constants, use a forward selection procedure to select an appropriate model for $\sigma$;
	
	\item Given the models for $\mu$ and $\sigma$ obtained in 1 and 2, respectively, and with $\tau$ fitted as constant, use a forward selection procedure to select an appropriate model for $\nu$;
	
	\item  Given the models for $\mu$, $\sigma$ and $\nu$ obtained in 1, 2 and 3, respectively, use a forward selection procedure to select an appropriate model for $\tau$;
	
	\item  Given the models for $\mu$, $\sigma$ and $\tau$ obtained in 1, 2 and 4, respectively, use a backward selection procedure, from the model for $\nu$ given by 3, to select an appropriate model for $\nu$;
	
	\item Given the models for $\mu$, $\nu$ and $\tau$ obtained in 1, 5 and 4, respectively, use a backward selection procedure, from the model for $\sigma$ given by 2, to select an appropriate model for $\sigma$;
	
	\item Given the models for $\sigma$, $\nu$ and $\tau$ obtained in 6, 5 and 4, respectively, use a backward selection procedure, from the model for $\mu$ given by 1, to select an appropriate model for $\mu$ and then stop.
\end{enumerate}

The model resulting from this selection procedure usually contains different subsets of terms (not necessarily the same) for each of the distribution's parameters. Once the best predictors for each model were established, the AIC and BIC \citep{Schwarz1978} for each model were obtained and Table \ref{criteria} shows the resulting comparison.
	
\begin{table}[ht!]
	\centering
	\caption{Zero-inflated models for the \textit{Trajan} data set obtained using the stepwise selection of covariates.}
	\label{criteria}      
	\begin{tabular}{lccccccc}
		\hline
		\multicolumn{1}{c}{Model} & -2$\times$logLik & AIC & BIC 		& $\mu$ 	  & $\sigma$ 	& $\nu$ 	  & $\tau$ \\ \hline
		ZIP               		  & 1251.613	& 1259.613 & 1274.007	& photoperiod & photoperiod & - 		  & -  \\
		ZINB                	  & 1233.623 	& 1245.623 & 1267.214 	& photoperiod & photoperiod & photoperiod & -  \\
		ZIPIG$^1$                 & 1241.760 	& 1251.760 & 1269.750	& photoperiod & intercept 	& photoperiod & -  \\
		ZIBNB$^1$     	  & 1238.955	& 1250.955 & 1272.545	& photoperiod & intercept 	& intercept   & photoperiod \\ \hline      
		\end{tabular}
\small $^1$The model was fit defining the predictor for each parameter rather than using stepwise selection.
\end{table}

As found by \citet{Ridout98}, the ZINB model is the one with the lowest AIC and BIC for the \textit{Trajan} data set. However, the criteria values shown for the ZIP and ZINB model in Table \ref{criteria} are lower than the values they found, because the deviance of GLMs and GAMLSSs are defined differently. The deviance of a GLM is defined as $D_{GLM}=-2 \log({\hat{L}_{c}}) + 2\log({\hat{L}_{s}})$, where $\hat{L}_{c}$ is the fitted likelihood of the current fitted model and $\hat{L}_{s}$ that of the saturated model (where a parameter is fitted for each observation in modeling $\mu$); while the GAMLSS's deviance is defined as $D_{GAMLSS}=-2 \log (\hat{L}_{c})$.

To diagnose the fitted ZINB GAMLSS, worm plots \citep{VanBuuren2001} and a classical Normal Q-Q plot of the normalized quantile residuals were produced and are depicted in Figures 4 and 5. Quantile residuals were introduced by \citet{Dunn96} and are used for GAMLSSs because they follow a standard normal distribution when the assumed model is correct whatever distribution is assumed for the response variable.

\begin{figure}[h]
	\begin{center}
		\begin{minipage}[t]{7cm}
		\includegraphics[width = 7cm, height = 6cm]{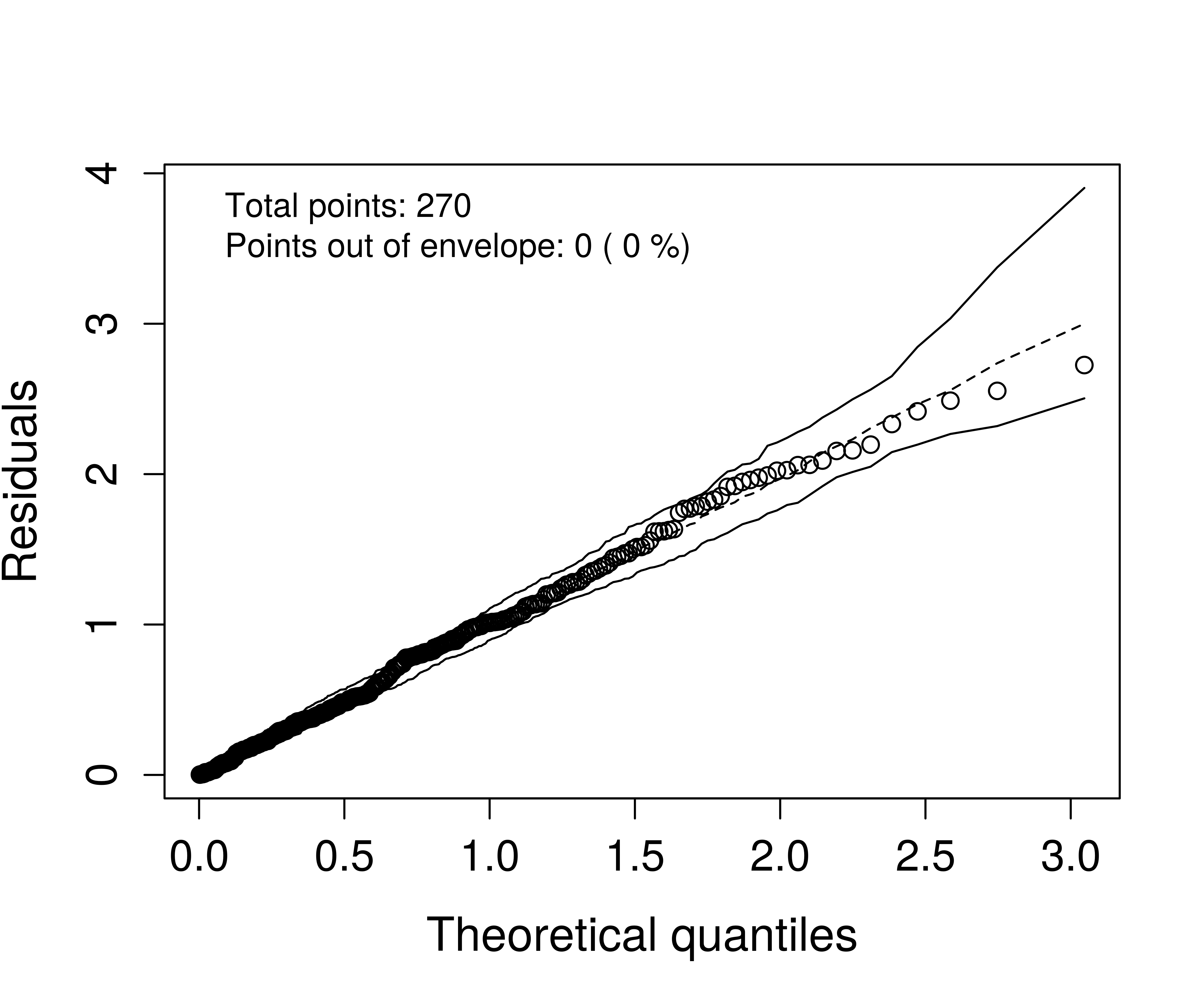}
		\end{minipage}
		\begin{minipage}[t]{7cm}
		\includegraphics[width = 7cm, height = 6cm]{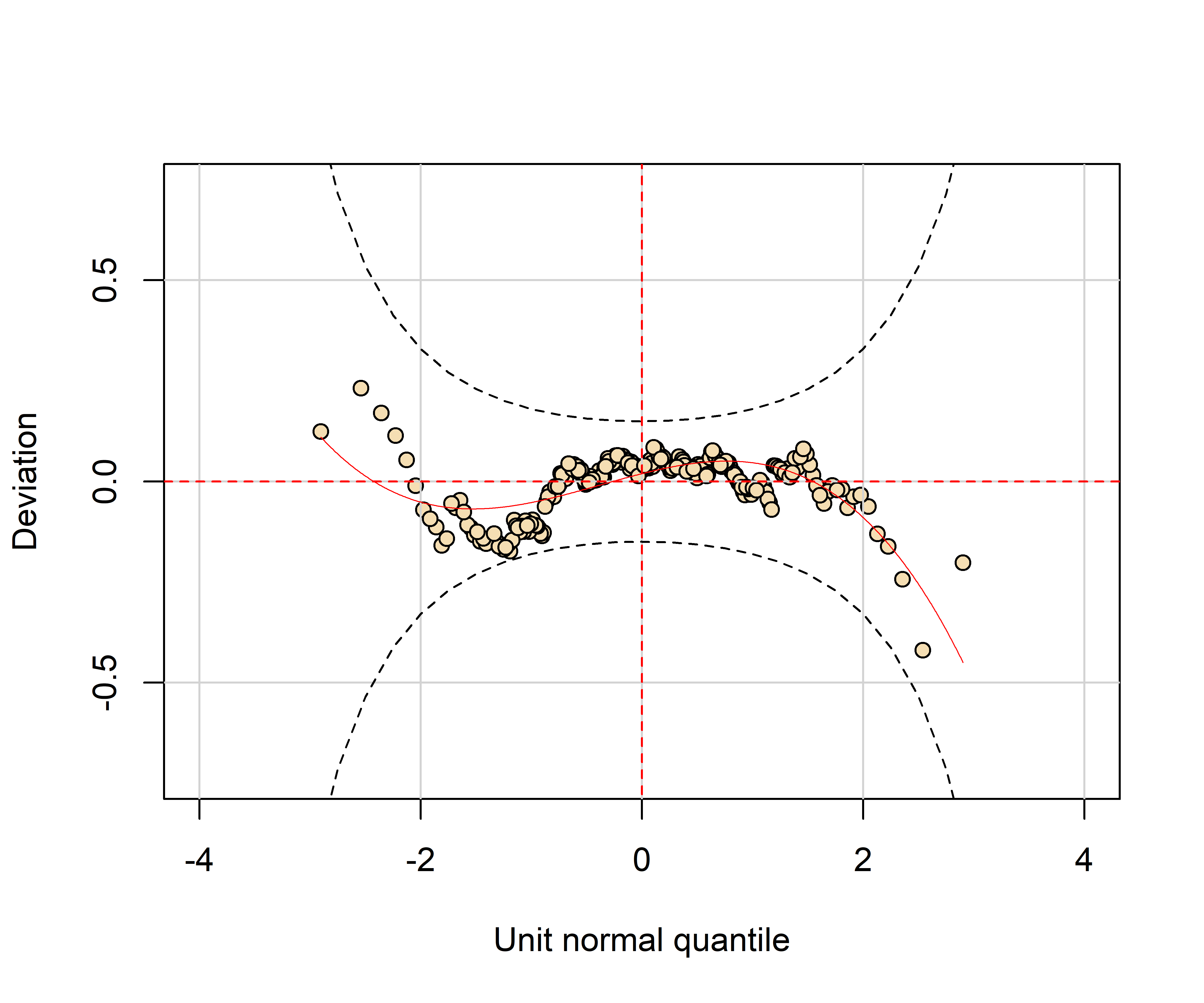}
		\end{minipage}
		\label{qq}
	\end{center}
	\caption{Normal Q-Q plot of the quantile residuals of the ZINB model produced with hnp package \citep{Moral2017} (left). Worm plot of the quantile residuals of the ZINB model, which is a detrended version of the normal Q-Q plot (right).}
\end{figure}

\begin{figure}[h]
	\begin{center}
			\includegraphics[width = 12cm, height = 8cm]{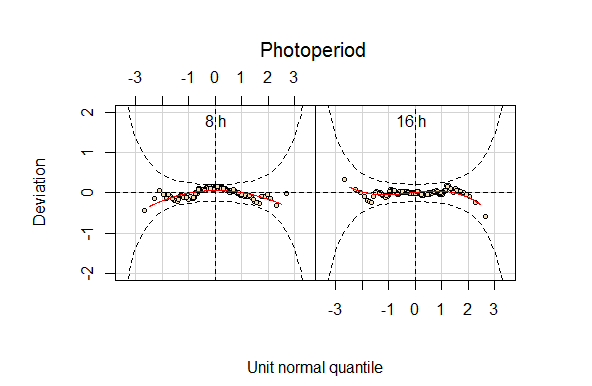}
		\label{wps}
	\end{center}
	\caption{Worm plots of the quantile residuals of the ZINB model related to each level of the photoperiod factor.}
\end{figure}

The right plot of Figure 4 shows a worm plot of the ZINB model's residuals. Worm plots are detrended Q-Q plots that highlight possible departures from the residuals' normality assumption. Figure 5 shows one worm plot for each level of the factor photoperiod duration. In this case failures of the model within each level of the explanatory variable are verified. From these two plots and also from the normal Q-Q plot (left plot of Figure 4) it can be seen that the residuals are within the 95\% confidence bands. In addiction, the residual shape lines are roughly horizontal in each of the worm plots and residual points are mostly on the Q-Q line for the normal Q-Q plot. Therefore, it is concluded that there is no visual departure from normality and that the residuals do not show systematic patterns. 

The fitted linear predictors found for each of the ZINB parameters are
\begin{equation}
\log(\hat{\mu})= 1.9725 \times \mbox{photo}_{8} +1.6954 \times \mbox{photo}_{16}, \\
\label{mu}
\end{equation}
\begin{equation}
\log(\hat{\sigma})= -3.1158 \times \mbox{photo}_{8} -1.8286 \times \mbox{photo}_{16} \\
\label{sigma}
\end{equation}
and
\begin{equation}
\mbox{logit}(\hat{\nu})= -4.3808 \times \mbox{photo}_{8} -0.1351 \times \mbox{photo}_{16}.
\label{nu}
\end{equation}

\begin{figure}[!htb]
	\centering
	\includegraphics[width=12cm, height=10cm]{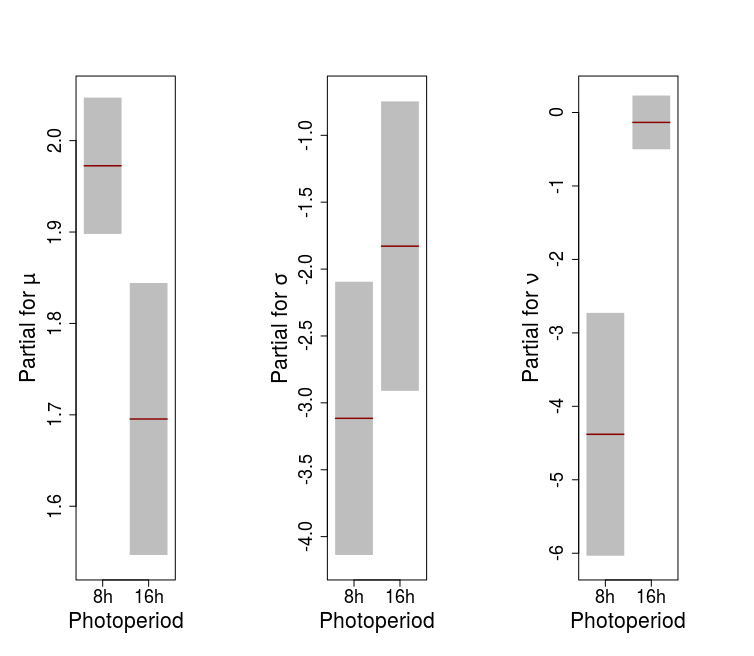}
	\caption{Termplots for the parameters $\mu$, $\sigma$ and $\nu$ of the ZINB distribution fitted.}
	\label{tp}
\end{figure}

To facilitate the interpretation of these predictors, the \verb|gamlss| package provides the function \verb|term.plot()|. This function produces plots of each covariate's estimates in the predictor of each distribution's parameter. Figure \ref{tp} shows the term plots related to the photoperiod level's estimates in predictors \ref{mu}, \ref{sigma} and \ref{nu} from left to right, respectively. The estimates in this figure are in the scale of the link function used. Point estimates are represented by the horizontal lines in Figure \ref{tp} and the shaded areas correspond to pointwise 95\% confidence intervals for the factor levels.
From the left term plot of Figure \ref{tp} and predictor (\ref{mu}) there is evidence of positive and different effects of the photoperiod durations in the mean number of roots, because both estimates are positive and their standard errors do not intersect nor contain the zero value. By the same reason, the term plot for $\sigma$ shows some evidence of overdispersion in the second group (since $\sigma$ represents the overdispersion parameter of the NB distribution), not significant though. This means that a model with only the intercept for the $\sigma$ parameter would give a similar fit. Finally, from the term plot for photoperiod in the predictor of the zero inflation parameter $\nu$ (on the right), it can be seen that the 16h photoperiod has a much higher effect than the lower duration, as expected once the zero inflation was observed mainly in that group. 
% From the idea suggested in the left plot, should not the right one have the 16h effect above the 0 line and the 8h one on it? Why was the 8h below and 16h on?

\section{Conclusion}
This work described the analysis of a zero-inflated count data set provenient from a real experiment. Using specific GAMLSSs' tools, the best response distributions were chosen and predictors for the parameters were obtained. Worm plots facilitated the residual diagnosis in the model globally and within ranges of the significant factor. The interpretation of the model's results was aided by term plots, which showed the effects of the covariate on each parameter of the response distribution. The best GAMLSS fitted for the number of roots was a ZINB model with the photoperiod duration as a significant factor for both the mean and the zero inflation parameters. This analysis showed that GAMLSSs are a competitive and flexible approach to model complex data structures, such as ones with excesses of zeros and overdispersion. Furthermore, the results obtained with this analysis agree with what was found by \citet{Ridout98} and support the conclusion of \citet{Marin93} that in \textit{Trajan} there is little effect of BAP concentration on the number of roots generated. 

\clearpage

%\begin{thebibliography}{99}
%
%\bibitem[Yoccoz, Nichols \& Boulinier(2001)]{Yoccoz}Yoccoz, N.G., Nichols, J.D. \& Boulinier, T. (2001) Monitoring of biological diversity in space and time. \textit{Trends in Ecology and Evolution}, 16, 446--453.
%
%\end{thebibliography}

\renewcommand\bibname{References} %Muda "Referência Bibliográficas" para Referências
\bibliographystyle{./genetics} %Localização do arquivo genetics.bst

\clearpage

\end{document}